# Predicting the chemical stability of monatomic chains

Zheng-Zhe Lin[1)†] and Xi Chen[2)]

1) *School of Science, Xidian University, Xi'an 710071, P. R. China*

2) *Department of Applied Physics, School of Science, Xi'an Jiaotong University, Xi'an 710049, P.R. China*

[†] Corresponding Author. E-mail address: linzhengzhe@hotmail.com

**Key Words:** Monatomic carbon chains, Stability of nanosystems

PACS **81.07.Vb** - Quantum wires

PACS **82.20.Db** - Transition state theory and statistical theories of rate constants

PACS **34.50.Lf** - Chemical reactions

**Abstract** - A simple model for evaluating the thermal atomic transfer rates in nanosystems [*EPL* **94**, 40002 (2011)] was developed to predict the chemical reaction rates of nanosystems with small gas molecules. The accuracy of the model was verified by MD simulations for molecular adsorption and desorption on a monatomic chain. By the prediction, a monatomic carbon chain should survive for $1.2 \times 10^2$ years in the ambient of 1 atm $O_2$ at room temperature, and it is very invulnerable to $N_2$, $H_2O$, $NO_2$, CO and $CO_2$, while a monatomic gold chain quickly ruptures in vacuum. It is worth noting that since the model can be easily applied via common *ab initio* calculations, it could be widely used in the prediction of chemical stability of nanosystems.

## I. Introduction

Since the birth of nanotechnology, preparation of thinner materials has gained great attention for their possible applications in emerging electronics, and many efforts were concentrated on finding stable one or two-dimensional nanocrystals. Over the past two decades, one-dimensional monatomic gold chains (MGCs) were prepared by pulling two contacted atom-sized junctions [1, 2]. Similar technique was used for the preparation of copper, aluminum and platinum chains [3]. Meanwhile, indirect evidence for the existence of one-dimensional monatomic carbon chains (MCCs) was found in the laser ablation of carbon nanotubes [4] or the condensation of carbon atomic gas [5]. In recent years, following the successful preparation of free-standing two-dimensional graphene crystals [6-9], free-standing MCCs were carved out from single-layer graphene by a high-energy electron beam [10], or unraveled from sharp carbon specimens [11, 12] or carbon nanotubes [13]. However, until now the stability of monatomic chains at room temperature is still unknown because *in situ* observations always make damages to them. For example, a MCC-graphene joint survives for about 100 s under irradiation of an electronic beam (4 A/cm$^2$ in density accelerated by a voltage of 120 kV) [10], or the body of a 10-atoms MGC survives for less than several seconds under the irradiation of a 30 A/cm$^2$ electronic beam [14, 15]. Since two-dimensional graphene has been proposed to be the material of next-generation circuit [16-21] with its remarkable electronic properties [22, 23], MCCs are expected to play a role of the thinnest natural wires in graphene-based circuits. Clearly, for the design of low-dimensional nanocircuits, the stability prediction of monatomic chains is highly desired to prejudge which low-dimensional nanodevices are stable at room temperature and deserve to be developed for practical applications.

More than 70 years ago, Landau and Peierls argued that low-dimensional crystals were thermodynamically unstable and could not exist [24, 25]. However, this theory was strongly challenged by the successful preparation of two-dimensional graphene [6-9]. Until now, we still do not have a powerful model to accurately predict the stability of low-dimensional crystals. Recently, a practical mechanical procedure was proposed to prepare long MCCs for the medium of tunable infrared laser [26] by unraveling single-layer graphene [27, 28], and so the stability prediction of MCCs is currently needed to guide relevant experiment exploration. Molecular dynamics (MD) simulation seems a direct approach to calculate the lifetime, i.e. the stability, but the timescale of MD cannot go beyond several microseconds. So, it is very necessary to build a uniform physical model for predicting the stability of nanosystems.

Recently, a statistical mechanical model was provided to predict the stability of nanosystems [29], which can be conveniently implemented via common *ab initio* calculations without empirical parameters and has been successfully applied on predicting the bond breaking rate of nanosystems constituted by MCC and graphene [29]. In this work, the model was extended to predict the chemical reaction rates between nanosystems and small molecules in gas-phase. The bond ruptures rates of monatomic chains caused by thermal motions or chemical reactions with small molecules were calculated at different temperatures. According to the results, MCCs should survive for $1.2 \times 10^2$ years in the ambient of 1 atm $O_2$ at room temperature, and shows very invulnerability to $N_2$, $H_2O$, $NO_2$, CO and $CO_2$ molecules, while MGCs quickly rupture in absolute vacuum due to thermal motions.

**II. Theoretical model**

In nanosystems, property changes or disintegrations may happen even via once atom transfer event. In such process, corresponding atomic transfer usually involves one or two "key atoms" in a potential valley crossing over a static barrier $E_0$. In most cases the atomic kinetic energy ($\sim k_B T$) at the valley bottom is significantly smaller than $E_0$, and the atom vibrates many times within the valley before crossing over the barrier. For the atoms bounded in condensed matters or molecules, the kinetic energy (KE) distribution is determined by $f(\varepsilon) = \sum_i f_i(\varepsilon) e^{-E_i/k_B T} / \sum_i e^{-E_i/k_B T}$, here $f_i(\varepsilon)$ is the KE distribution of quantum state $E_i$, including all of the translational, rotational and vibrational states. As an example, $f(\varepsilon)$ of an individual atom in a $Cl_2$ molecule is shown in Fig. 1(a). At room temperature or above, the quantum state density of atoms approaches to continuum and the distribution $f(\varepsilon)$ turns into the classical one. In solid materials, the atomic motions are even more classical due to an amount of near continual vibrational states. In the classical limit, the Boltzmann KE distribution $\varepsilon^{1/2} e^{-\varepsilon/k_B T}$ for individual atoms can be easily derived from classical ensemble theory. For a classical mechanical system including $N$ atoms, the total energy $E = \vec{p}_1^2/2m_1 + ... + \vec{p}_N^2/2m_N + V(\vec{x}_1,...,\vec{x}_N)$ and the KE distribution of the i[th] atom reads

$$f(\varepsilon) = \int \delta[\vec{p}_i^2/2m_i - \varepsilon] e^{-E/k_BT} d\vec{p}_1...d\vec{p}_N d\vec{x}_1...d\vec{x}_N / \int e^{-E/k_BT} d\vec{p}_1...d\vec{p}_N d\vec{x}_1...d\vec{x}_N$$
$$= \int \delta[\vec{p}_i^2/2m_i - \varepsilon] e^{-\vec{p}_i^2/2m_ik_BT} d\vec{p}_i / \int e^{-\vec{p}_i^2/2m_ik_BT} d\vec{p}_i \quad , \quad (1)$$
$$= \frac{\varepsilon^{1/2} e^{-\varepsilon/k_BT}}{\sqrt{\pi}(kT)^{3/2}/2}$$

holding for atoms in any condensed matter or molecule. This distribution is in very good agreement with various MD simulations, and it was proved that the ergodicity is achieved in a time less than 100 ps at room temperature or above [29]. By this distribution, the atomic probability for having a KE $\varepsilon$ larger than $E_0$ is

$$P = \frac{\int_{E_0}^{+\infty} \varepsilon^{1/2} e^{-\varepsilon/k_BT} d\varepsilon}{\int_0^{+\infty} \varepsilon^{1/2} e^{-\varepsilon/k_BT} d\varepsilon} \quad . \quad (2)$$
$$= \frac{\int_{E_0}^{+\infty} \varepsilon^{1/2} e^{-\varepsilon/k_BT} d\varepsilon}{\sqrt{\pi}(k_BT)^{3/2}/2}$$

With a vibration frequency $\Gamma_0$, the atomic transfer rate over the barrier reads [29]

$$\Gamma = \Gamma_0 \frac{\int_{E_0}^{+\infty} \varepsilon^{1/2} e^{-\varepsilon/k_BT} d\varepsilon}{\sqrt{\pi}(k_BT)^{3/2}/2} . \quad (3)$$

For a given $\varepsilon$ at the valley bottom, the oscillation period $\tau(\varepsilon) = \sqrt{m}\int d\bar{x}/2[\varepsilon - V(\bar{x})]$ along the minimum energy path (MEP) [29] can be determined by the potential $V(\bar{x}) = \int \vec{F}(\bar{x}) \cdot d\bar{x}$, where $\vec{F}(\bar{x})$ is the force felt by the key atom at position $\bar{x}$. With the corresponding oscillation frequency $v(\varepsilon)=1/\tau(\varepsilon)$, the averaged frequency reads [29]

$$\Gamma_0 = \frac{\int_0^{E_0} v(\varepsilon)\varepsilon^{1/2} e^{-\varepsilon/k_BT} d\varepsilon}{\int_0^{E_0} \varepsilon^{1/2} e^{-\varepsilon/k_BT} d\varepsilon} . \quad (4)$$

It is worth noting that the $\Gamma_0(T)$ given by Eq. (4) is in good agreement with the value observed in MD simulations [29]. For transfers involving two key atoms, the event occurs when the KE sum $\varepsilon_1+\varepsilon_2$ of key atoms is larger than $E_0$, and therefore the corresponding rate should be

$$\Gamma = \Gamma_0 \frac{\int_{\varepsilon_1+\varepsilon_2 \geq E_0} \varepsilon_1^{1/2} \varepsilon_2^{1/2} e^{-(\varepsilon_1+\varepsilon_2)/k_B T} d\varepsilon_1 d\varepsilon_2}{\int_{\varepsilon_1 \geq 0, \varepsilon_2 \geq 0} \varepsilon_1^{1/2} \varepsilon_2^{1/2} e^{-(\varepsilon_1+\varepsilon_2)/k_B T} d\varepsilon_1 d\varepsilon_2}$$

$$= \Gamma_0 \frac{\int_{E_0}^{+\infty} \varepsilon^2 e^{-\varepsilon/k_B T} d\varepsilon}{2(k_B T)^3} . \quad (5)$$

In our previous work [29], the above model has been verified by MD simulation and successfully applied to predict the stability of MCC-graphene joint and carbon-carbon bonds in MCC body, reproducing results in good agreement with the experimental data and showing an accuracy better than the conventional transition state theory.

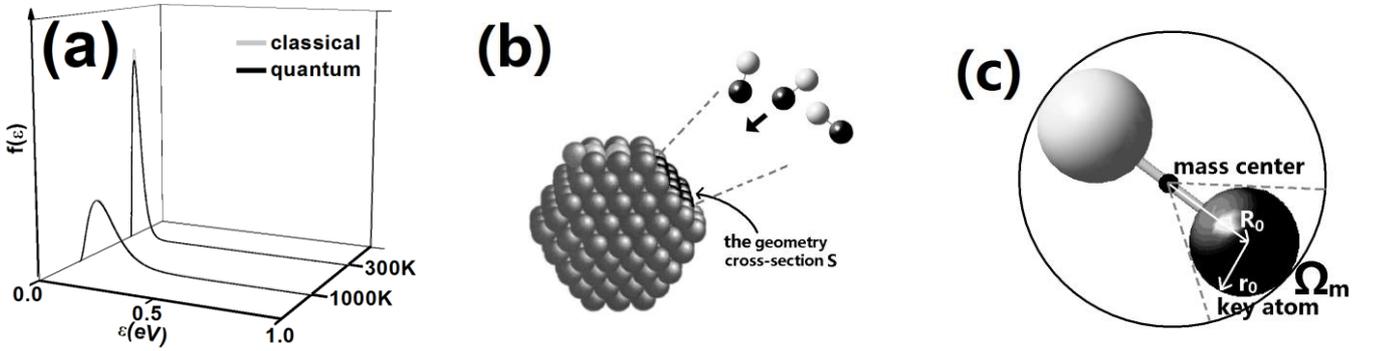

Fig.1 The KE distribution $f(\varepsilon)$ of an atom in a $Cl_2$ molecule by classical (gray lines) and quantum mechanics (black lines) at 300 and 1000 K (a). The geometry cross-section $S$ of the key atoms in the nanosystem (b) and the solid angle of the key atom opened in a molecule (c).

For chemical reactions of nanosystems with small gas molecules, an atomic event takes place when the incident molecule hits the key atoms in the nanosystem with a specific orientation and a translational KE $\varepsilon$ larger than $E_0$ [Fig. 1(b)]. By the classical ensemble theory, the translational KE distribution of molecular mass center is also Boltzmann. So, for reactant molecules at a concentration $c$, the reaction rate reads

$$\Gamma = \frac{\sigma v c}{2} \frac{\int_{E_0}^{+\infty} \varepsilon^{1/2} e^{-\varepsilon/k_B T} d\varepsilon}{\sqrt{\pi}(k_B T)^{3/2}/2}, \quad (6)$$

where $\sigma$ is the effective cross-section of the nanosystem and $v = \sqrt{2k_B T/\pi M}$ is the average molecular thermal velocity along the cross-section normal. The factor 2 in the denominator is because only half of the molecules move towards the cross-section. It should be noted that $\sigma$ is not equal to the geometry cross-section $S$ of the key atoms in the nanosystem [Fig. 1(b)], but instead $\sigma = S\Omega_m/4\pi$, where $\Omega_m = 2\pi(1-\sqrt{1-r_0^2/R_0^2})$ is the solid angle opened by the molecular key atoms [Fig. 1(c)], with $R_0$ the atomic distance to the molecular mass center and $r_0$ the atomic radius. So, the effective cross-section reads

$$\sigma = S(1-\sqrt{1-r_0^2/R_0^2})/2. \quad (7)$$

In practical applications, $r_0$ can be simply taken as the atomic covalent radius [30].

### III. MD simulations

In this section, the applicability of the model to molecular adsorption and desorption reactions on a monatomic chain was verified by MD simulations. In a periodic cubic box with a side length of 30 Å, the simulation system was set up by putting a 20-atom MCC and a diatomic molecule along with 33 helium atoms as the buffer gas (BG). The terminal atoms of MCC were set fixed, and the pressure of BG is about 50 atm at 300 K. Simulations for the adsorption was initialized by putting the diatomic molecule in a random position, and the adsorption takes place when the molecule clings to the MCC (the upper sketch in Fig. 2(a)). For the desorption, the molecule was initialized on the MCC and then goes away (the lower sketch in Fig. 2(a)). The interaction between carbon atoms is described by the Brenner potential [31, 32], and Leonard-Jones potential is applied for carbon-BG and BG-BG interactions [33]. For the molecule, the interaction between its two atoms reads

$$V_{mm}(r) = C_1 e^{-C_2 r} - C_3 e^{-C_4 r} \quad (8)$$

with a bond energy of 1.05 eV ($C_1=9.073\times10^5$ eV, $C_2=10.925$ Å$^{-1}$, $C_3=3.514$ eV, $C_4=0.764$ Å$^{-1}$). In order to provide a barrier for the molecular adsorption and desorption progress, a modified Leonard-Jones potential

$$V_{cm}(r) = D_1/r^{12} - D_2/r^6 + D_3/r^3 \quad (9)$$

is designed for the interaction between carbon atoms and the molecular atoms ($D_1=3.028\times10^3$ eV, $D_2=3.177\times10^2$ eV, $D_3=33.348$ eV). These parameters for the artificially constructed potential, i.e. Eq. (8) and (9), were adjusted to let the adsorption and desorption happen within the time scale of MD simulations. Because our model does not depend on the specific form of interaction potential, it is suitable for the adsorption and desorption progress of any diatomic molecule on a monatomic chain, and the chosen parameters for Eq. (8) and (9) do not affect the verification of the model. Simulations were initialized at a given temperature $T$, and the thermal motion of BG was controlled by a thermal bath which randomly chooses an atom $i$ and replaces its velocity $v_i^{old}$ with $v_i^{new}$ in a time interval [34]. Here,

$$v_i^{new} = (1-\theta)^{1/2} v_i^{old} + \theta^{1/2} v_i^T \quad (i=x,y,z), \tag{10}$$

where $v_i^T$ is a random velocity chosen from the Maxwellian distribution and $\theta=0.1$ [35] is a random parameter controlling the strength of velocity reset. By our FORTRAN code based on the velocity Verlet algorithm and a time step of 0.2 fs, MD simulations were performed repeatedly at every temperature point in the range of 700~2000 K until the change of average reaction rate $\Gamma$ was below 5%.

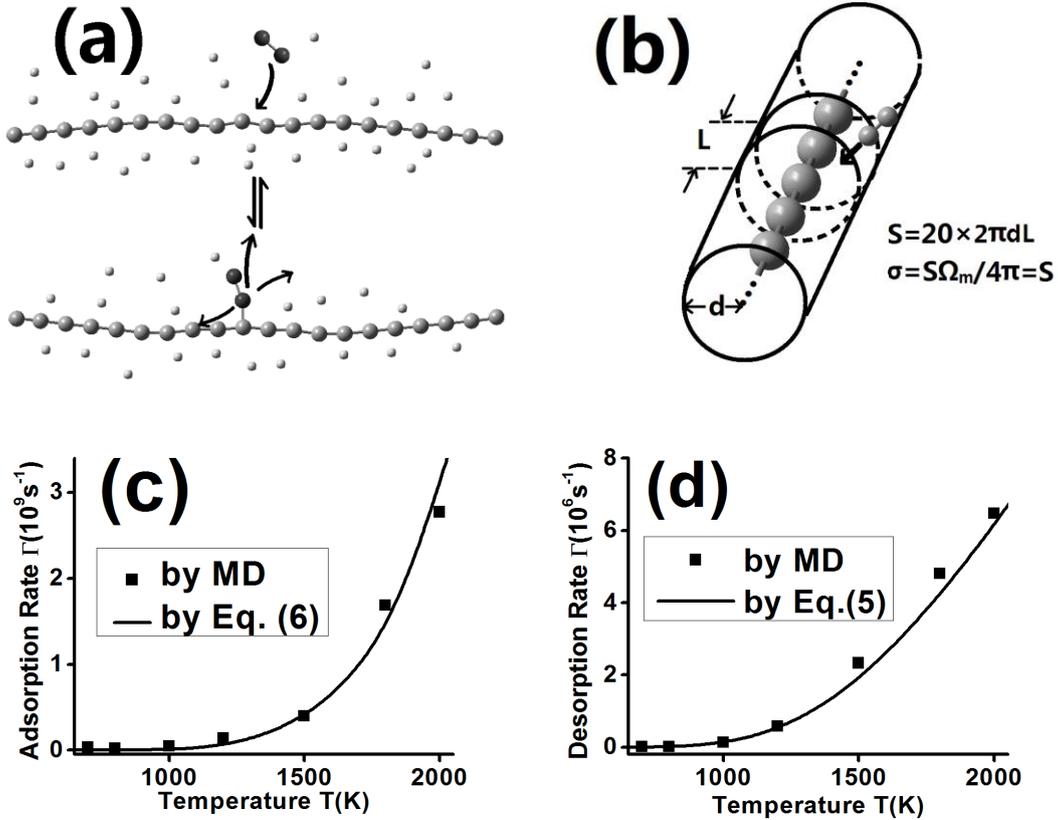

Fig. 2. The simulation system for molecular adsorption and desorption on a MCC (a); the cross-section $\sigma$ for the adsorption (b); the adsorption (c) and desorption (d) rates via MD simulations and the model.

The MEPs of the adsorption and desorption were calculated using the pseudo reaction coordinate method [36], recognizing the corresponding barriers $E_{0a}$=1.052 eV and $E_{0d}$=0.837 eV, respectively. The cross-section $\sigma$ was estimated as follows. The geometry cross-section of the 20-atom MCC is $S = 20 \times 2\pi dL = 1.10 \times 10^3$ Å$^2$, with $d$= 6.74 Å the distance from the MCC axis to the molecule mass center where the barrier $E_{0a}$ appears and $L$=1.30 Å the average carbon-carbon bond length in the MCC [Fig. 2(b)]. By the interaction potential Eq. (9), the solid angle taken by an atom in the molecule gets close to $2\pi$ because the atomic radius $r_0$ is close to its distance to molecular mass center $R_0$. So, for the sum of two atoms $\Omega_m = 4\pi$, and the cross-section of the 20-atom MCC is $\sigma = S\Omega_m/4\pi = S$. According to the results, the molecular adsorption rates Γ predicted by Eq. (6) are in good agreement with MD results [Fig. 2(c)]. It is worth noting that the model is also applicable to triatomic or polyatomic molecules because Eq. (6) is independent of the molecular geometry.

The desorption progress happens when the bond between the MCC and the molecule breaks, i.e. the KE sum $\varepsilon_1+\varepsilon_2$ of the two atoms is larger than $E_{0d}$ [Eq. (5)]. The molecule can go away from the MCC along the radial and two tangential directions (the lower sketch of Fig. 2(a)), and so, the calculated rate Γ should be multiplied by 3. Indeed, three equivalent paths were found in the MEP calculations. For triatomic or polyatomic molecules, Eq. (5) is also applicable because it only concerns the two atoms of the bond. The oscillation frequency $\Gamma_0$ was evaluated as $3.6 \times 10^{12}$~$4.0 \times 10^{12}$ s$^{-1}$ in the simulation temperature range. According to the results, the desorption rates Γ calculated by Eq. (5) are in good agreement with the MD results [Fig. 2(d)] as well as that for the adsorption, showing the accuracy of our model for chemical reactions of MCCs with small molecules.

## IV. Application

To study the stability of monatomic chains, the rate of thermal bond ruptures in MCCs and MGCs and chemical reactions of MCCs with common $N_2$, $O_2$, $H_2O$, $NO_2$, CO and $CO_2$ molecule in the air were investigated. To apply the model, the geometry optimization, reaction barriers $E_0$, MEPs and the forces $F(\bar{x})$ felt by the key atom were investigated for a 30-atom MCC and MGC with their terminals fixed. All the calculations were performed on level of density functional theory (DFT) via the Gaussian 03 package [37] with the newly developed hybrid X3LYP functional [38] which is considered more accurate than other functionals in the potential surface and MEP calculations. The 6-31G(d,p) basic set were employed, except using LanL2DZ basic set for gold atoms. Canonical modes for the geometries of potential minima and transition states were calculated to confirm the results. To verify the calculation technique, the adsorption geometry and energy of $NO_2$ on graphene sheet were investigated, finding an adsorption energy of 0.056 eV which is close to the result via Perdew-Burke-Ernzerhof (PBE) functional [39].

The thermal ruptures in monatomic chains shown in Fig. 3(a) are attributed to the motions of neighboring atoms in opposite directions [29]. Such motions can be decomposed in two independent directions perpendicular to the chain axis. Indeed, two equivalent MEPs were found in the calculations. For MCCs, the rupture barrier $E_0$=4.96 eV is close to the value via PBE functional [29]. By Eq. (5), at 300 K the lifetime $\tau=1/\Gamma$ of a carbon-carbon bond in the MCC is about $2\times10^{58}$ years [Fig. 3(b)], and a MCC of 1 cm in length (about $8\times10^7$ bonds) should survive for $3\times10^{50}$ years, indicating that MCCs are very stable in vacuum at room temperature. Even at 1000 K, a carbon-carbon bond in the MCC of 1 cm should survive for about 11 years [Fig. 3(b)]. It should be noted that no body ruptures of MCC were observed in experiments [10], and a long-living MCC has been prepared by some scientists [11]. With $E_0$=1.37 eV, the lifetime of an 10-atom MGC should be 10 days at 300 K, and sharply declines to 3 s at 400 K [Fig. 3(b)], which is quite close to the experimental results [15].

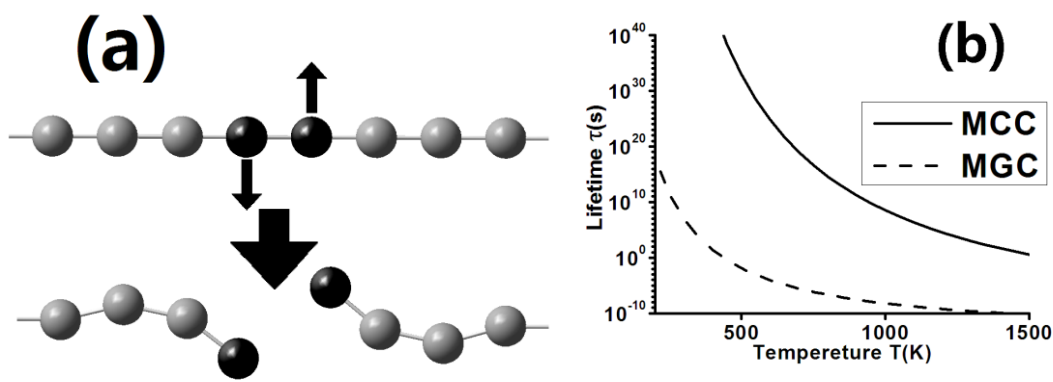

Fig. 3. Bond rupture in a monatomic chain (a) and the lifetime of a bond in MCCs and MGCs at 200~1500 K (b).

Very weak interactions were found between the MCC and $N_2$, $H_2O$, $NO_2$, CO and $CO_2$ molecules. Along the MEPs of $NO_2$, CO and $CO_2$ molecules approaching the MCC, the potential drops to a valley of 0.017~0.024 eV without barriers, while only repulsive interactions were found for $N_2$ and $H_2O$. For the lowest energy configuration, little deformation of the MCC was found, and the balance distance of $NO_2$, CO and $CO_2$ molecules to the MCC axis is about 3.2~3.6 Å. For these barrierless adsorptions, Eq. (6) becomes $\Gamma_a = \sigma v c/2$. By $\sigma \approx 28.0$ Å$^2$, the adsorption rate of 1 atm $NO_2$ on one carbon atom is about $\Gamma_a = 6.4 \times 10^8$ s$^{-1}$ at 300 K, and corresponding desorption rate [Eq. (5)] was estimated to be $\Gamma_d = 2.8 \times 10^{12}$ s$^{-1}$. So, at 300 K, the molecular coverage ratio of a MCC in the ambient of 1 atm $NO_2$ is $R = \Gamma_a/(\Gamma_a + \Gamma_d) \approx 0.02$ %. At 1000 K, the ratio even decreases to $R \approx 0.01$ %. Similar situations are also found for CO and $CO_2$. Such weak adsorption or even repulsion means that the molecules can hardly break the carbon-carbon bonds of the MCC, presenting the chemical invulnerability of MCCs to these molecules.

In the calculation of MEP, a two-step process was found for the reaction of the MCC with $O_2$ molecules. Firstly, an $O_2$ molecule approaches the MCC and turns into the adsorption configuration A in Fig. 4(a) with one oxygen atom bonds with a carbon atom. Secondly, the other oxygen atom gets close to the carbon atom neighboring to the newly formed bond and becomes the configuration B in Fig. 4(a), and then transfers to the other side. If all the carbon atoms adsorb oxygen molecules and change into the configuration B, the whole MCC will disintegrate into many carbon oxide molecules, i.e. the configuration C in Fig. 4(a). In the first molecular adsorption step [Eq. (6)], the system climbs over a barrier $E_0^{1st+}$=0.93 eV and reaches the configuration A with decreasing carbon-oxygen bond length [Fig. 4(b)], and the corresponding inverse bond-breaking progress [Eq. (5)] has a barrier $E_0^{1st-}$=0.38 eV [Fig. 4(b)]. In the second step [Eq. (3)], the configuration B forms with the decreasing length of the other carbon-oxygen bond after climbing over a barrier $E_0^{2nd+}$=0.49 eV [Fig. 4(c)]. For the corresponding inverse progress [Eq. (3)], since the barrier $E_0^{2nd-}$=3.55 eV is much higher than $E_0^{2nd+}$ [Fig. 4(c)] the rate is much slower than the forward one in about 50 orders of magnitude. The total oxidation rate of the MCC can be estimated by the kinetic equations. Note $N_A$ and $N_B$ as carbon atoms in the configuration A and B, respectively, and $N$ the total carbon atoms in the MCC. As an intermediate state, the steady-state equation of the configuration A reads

$$0 = dN_A/dt = \Gamma_{1st+}(N - N_A - N_B) - (\Gamma_{1st-} + \Gamma_{2st+})N_A, \tag{10}$$

and the total oxidation rate should be

$$\begin{aligned} dN_B/dt &= 2\Gamma_{2st+}N_A \\ &= 2\Gamma_{1st+}\Gamma_{2st+}(N-N_B)/(\Gamma_{1st+} + \Gamma_{1st-} + \Gamma_{2st+}) \\ &\approx 2\Gamma_{1st+}\Gamma_{2st+}N/(\Gamma_{1st+} + \Gamma_{1st-} + \Gamma_{2st+}) \end{aligned} \tag{11}$$

Then, the oxidation time of a whole MCC could be estimated by

$$\tau = N/(dN_B/dt) \approx (\Gamma_{1st+} + \Gamma_{1st-} + \Gamma_{2st+})/2\Gamma_{1st+}\Gamma_{2st+}. \tag{12}$$

At 300 K, a MCC in the ambient of 1 atm $O_2$ gas should survive for $1.2\times10^2$ years [Fig. 4(d)]. Even at 1000 K, the MCC should survive for 2 hours [Fig. 4(d)], indicating that MCCs are invulnerable to $O_2$ gas. So, MCCs should be very stable medium for tunable infrared laser [26] because they are more invulnerable in high vacuum (~$10^{-7}$ Pa of $O_2$).

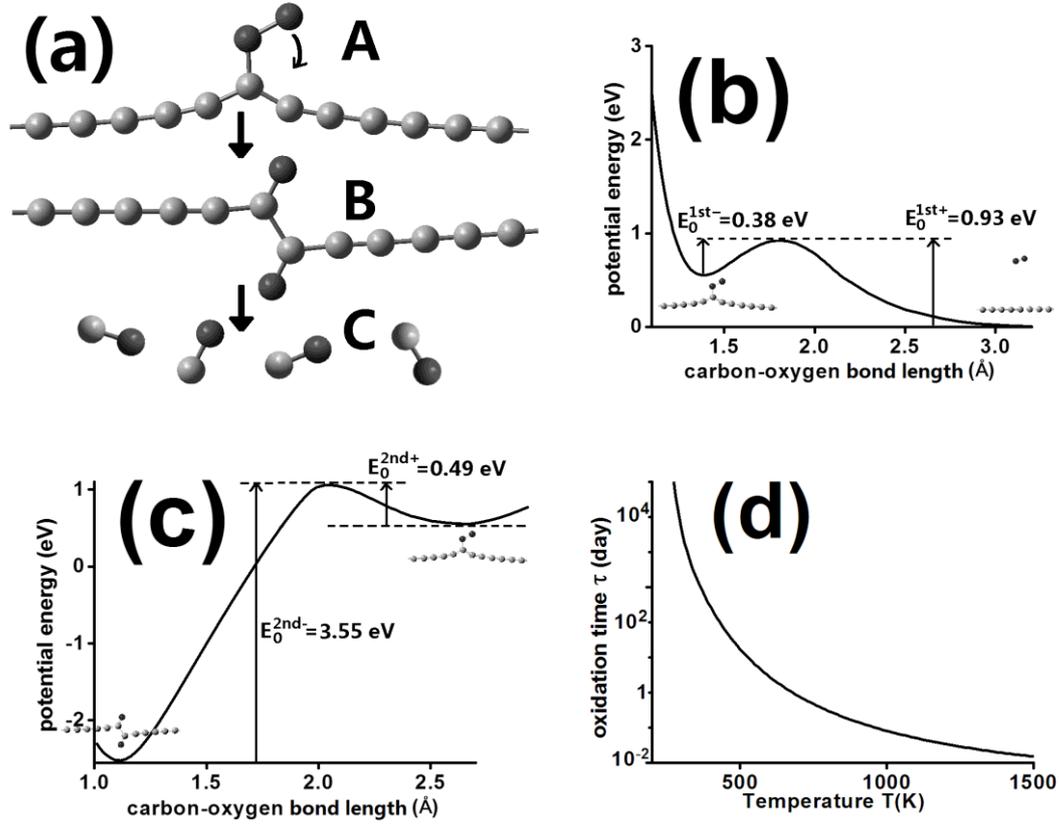

Fig. 4 Steps of the chemical reaction between a MCC and $O_2$ molecules (a). Corresponding potential profiles along the MEP are shown for the first (b) and second (c) reaction step, and the total oxidation time of a MCC (d) in the ambient of 1 atm $O_2$ was plotted under different temperatures.

**V. Summary**

In summary, a statistical mechanical model [29] was extended to predict the chemical reaction rates of nanosystems with small gas molecules. The model is based on the fact that the KE distribution of atoms or molecules always obeys $\varepsilon^{1/2} e^{-\varepsilon/k_B T}$, and the accuracy of the model has been verified by MD simulations. By the prediction, MCCs are very invulnerable to $N_2$, $O_2$, $H_2O$, $NO_2$, CO and $CO_2$ ambient at room temperature or above, while MGCs quickly rupture in thermal motions. This result reproduces the experiment data and suggests that short MCCs are good candidate for tunable laser medium [26]. Since our model needs only the static potential profile along the MEP, which can be easily obtained via common *ab initio* calculations, the new model could be widely used in the prediction of physical and chemical stability of nanosystems.

**Acknowledgements**

This work was supported by the Fundamental Research Funds for the Central Universities.